\documentclass[12pt,twoside,fleqn]{article}
\usepackage{epsfig}
\usepackage{psfig}
\def\lsim{\raise0.3ex\hbox{$<$\kern-0.75em\raise-1.1ex\hbox{$\sim$}}}
\def\gsim{\raise0.3ex\hbox{$>$\kern-0.75em\raise-1.1ex\hbox{$\sim$}}}
\makeatletter
\@addtoreset{equation}{section}
\makeatother

\newcommand{\beqn} {\begin{equation}}
\newcommand{\eqn} {\end{equation}}

\newcommand{\slsh}[1] {#1\kern-.43em/}
\newcommand{\real}{{\sf I}\kern-.12em{\sf R}}
\newcommand{\comp}{{\sf I}\kern-.48em{\sf C}}
\newcommand{\nin} {\in\kern-.6em/}

\def\MEF{m_{\rm eff}}\def\mef{\ifmmode\MEF\else$\MEF$\fi}

\begin{document}
\thispagestyle{empty}
%
 \mbox{} \hfill BI-TP 95/38\\
 \mbox{} \hfill December 1995\\
\begin{center}
{{\large \bf Lattice Regularized QCD at Finite Temperature\footnote{Lectures
given at the "Enrico Fermi School" on {\it Selected Topics in non
perturbative QCD}, \\
27 June - 7 July, Varenna, Italy} }
 } \\
\vspace*{1.0cm}
{\large Frithjof Karsch} \\
\vspace*{1.0cm}
{\normalsize
$\mbox{}$ {Fakult\"at f\"ur Physik, Universit\"at Bielefeld,
D-33615 Bielefeld, Germany}}\\
\end{center}
\setlength{\baselineskip}{1.3\baselineskip}

\baselineskip 10pt

\noindent
During the first part of this
review we will focus on the thermodynamics of $SU(N)$ gauge
theories at finite temperature. We will present results from a
calculation of electric and magnetic screening masses for the gluons,
discuss calculations of the critical temperature in units of the string
tension and bulk thermodynamic quantities like the energy density and
pressure. In particular, the latter calculations have now reached a stage
where $O(a^2)$ cut-off effects can be controlled systematically and an
extrapolation to the continuum limit can be performed.
In the second part we discuss the critical behaviour of QCD with light
quarks. We analyze the chiral transition in 2-flavour QCD and
present results on the temperature dependence of hadron properties.
\vskip 20pt

\baselineskip 15pt

\noindent
{\large \bf I. \hskip 0.3truecm Introduction}
\vskip 5pt

\noindent
Lattice studies of the thermodynamics of QCD have been
performed since more than ten years \cite{first}. 
The goal of these investigations is to understand quantitatively
the behaviour of strongly interacting matter at high temperatures. In 
order to achieve this computer simulations of lattice regularized QCD 
have to be performed with
parameters for the quark masses and number of flavours which are as close
as possible to the values in the {\it real world}, ie. a world where the
dynamics is controlled by 
two light, nearly massless quark flavours. In addition these
calculations should be performed as close as possible to the continuum 
limit in order to minimize systematic errors resulting from the
finite lattice cut-off. Numerical calculations are steadily improving in both 
respects.
Nonetheless, given the presently available computational resources,
one still has to compromise either by performing calculations in the physically
interesting regime with light quarks but relatively large lattice spacings
and limited statistical accuracy or by analyzing the approach to the 
continuum limit through high statistics calculations within the context 
of quenched QCD, ie. in pure $SU(N)$ gauge theories. 

Most of the results with high statistical accuracy have been obtained 
for pure $SU(N)$ gauge theories, which do already 
provide a highly non-trivial model for the thermodynamics of QCD. Many
of the global features of the thermodynamics like, for instance, 
the rapid rise of the energy density in the plasma phase 
towards the high temperature ideal gas limit, have been found to be
qualitatively similar in $SU(N)$ gauge theories and QCD with light 
quarks. Moreover, it is well known that 
also the {\it finite cut-off effects}, which are quite large in 
calculations of bulk thermodynamic observables, are similar in QCD and 
$SU(N)$ gauge theories. The simpler $SU(N)$ gauge
theory thus is an ideal model for the systematic investigation of
discretization errors and their controlled removal in the continuum limit.
We will discuss this in section II of these lectures.

The investigations of the QCD thermodynamics in the presence of light 
quarks, on the other hand, are not yet on the same quantitative level. The
analysis of the equation of state, for instance, still suffers from large 
discretization errors. Another important issue in the analysis of
the thermodynamics with two light quarks is to understand the role of
chiral symmetry restoration at the critical temperature. As this is an
infrared problem, the discretization errors do not play a central role in this
case. Here the main objectives are to model the correct continuum flavour
symmetry in lattice regularized QCD and to perform calculations at
small enough quark masses in order to be able to
observe the influence of chiral symmetry breaking/restoration on the QCD
thermodynamics. We will discuss this aspect in section III of these
lectures.

In this short presentation we will not be able to review systematically the 
formulation of lattice regularized QCD. For a more detailed discussion of
the basic concepts with emphasis put on the finite temperature formulation
we refer to recent reviews and references given therein
\cite{Kar90,DeT95,Kan95}.

\vskip 20pt
\noindent
{\large \bf II \hskip 0.3truecm Equation of State for SU(N) Gauge Theories}
\vskip 5pt

\noindent
{\large \bf II.1 \hskip 0.3truecm Screening masses and perturbation theory}
\vskip 5pt

\noindent
Our intuitive understanding of the high temperature phase of QCD and as such
also of the non-abelian $SU(N)$ gauge theories has been guided to a large
extent by perturbative calculations at high temperatures. Asymptotic freedom
suggests that at very high temperatures the QCD plasma approaches an ideal
gas of non-interacting quarks and gluons. This requires that
the temperature dependent running coupling, $g(T)$, becomes small.
At the same time, it
is well-known that perturbative calculations are faced with severe infrared
problems \cite{Linde} which only can be overcome, if thermal electric and magnetic
masses of order $g(T) T$ and $g^2(T) T$, respectively,
are generated for the gluons. Even then the perturbative expansion of
the thermodynamic potential is, however, known to show quite poor
convergence properties also at temperatures which are several times as large
as the critical temperature \cite{Arnold}. 
One thus may wonder to what
extent the concept of magnetic and electric screening masses, in particular
the arguments given for their relative ordering, $m_e (T)\simeq O(g(T)T) > 
m_m (T) \simeq O(g^2(T)T)$, are of relevance beyond perturbation theory.
After all, such arguments rely on the assumption that $g(T)$ will be small 
at high temperature.

\begin{figure}[htb]
\centerline{
   \epsfig{bbllx=50,bblly=50,bburx=310,bbury=240,
       file=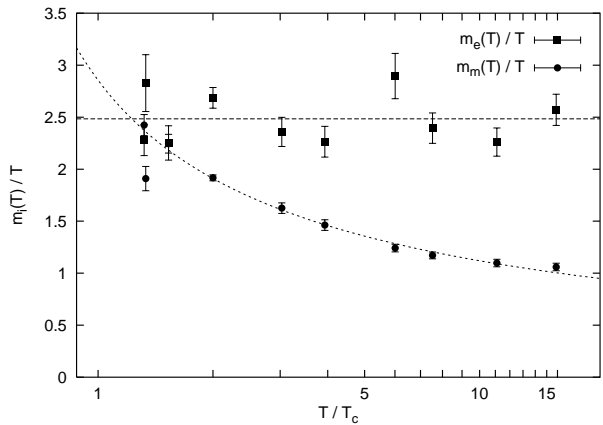,width=80mm}}
\caption{\baselineskip 10pt
Electric and magnetic gluon screening masses for the $SU(2)$ gauge theory
calculated in Landau gauge [5].
For the electric mass we show a straight line fit, $m_e = (2.484 \pm 0.052)
T$. For the magnetic mass we include the observed $g^2$-dependence,
$m_m = ( 0.466  \pm 0.015) g^2(T) T$, with $g^{-2} (T) = (11/12\pi^2)
\ln((3.8\pm 0.3)T/T_c)$}

\label{fig:screen}
\end{figure}

A recent analysis of the gluon propagator in a $SU(2)$ gauge 
theory \cite{Hel95}, performed in Landau gauge, shows that the expected 
ordering of magnetic and electric screening masses indeed persists down to
temperatures close to $T_c$, ie. $m_e (T) > m_m (T)$. However, it also
shows that at least the leading order perturbative relations for the
electric mass do not hold for temperatures at least up to $15T_c$. Rather
than decreasing with temperature $m_e(T)/T$ does stay
constant and is quite large, i.e. $m_e(T)/T \simeq 2.5$.
Although the magnetic mass does seem to scale like $m_m \simeq g^2(T) T$
and is smaller than $m_e$ above $T_c$, also in this case the running coupling 
extracted from such a scaling ansatz turns out to be large, 
i.e. $g^2(2T_c) \simeq 4$.
Results from a calculation in Landau gauge \cite{Hel95} are shown in
Figure~\ref{fig:screen}.

Although the gauge dependence of the results presented above still has to be
analyzed carefully, they do suggest that the running coupling
constant is larger than unity for temperatures up to a few times $T_c$.
This observation is also in agreement with conclusions drawn from the temperature
dependence of other observables like, for instance, the spatial string 
tension \cite{Bali93,Kar95}. One thus
cannot expect that perturbation theory provides a quantitative description
of bulk thermodynamics in this temperature regime \cite{Arnold}.
A non-perturbative approach is needed.

\vskip 20pt
\noindent
{\large \bf II.2 \hskip 0.3truecm Cut-off dependence of thermodynamics}
\vskip 5pt

\noindent
Lattice calculations of the energy density ($\epsilon$), pressure ($p$) and
other thermodynamic
variables led to some understanding of the temperature dependence of these
quantities in the QCD plasma phase. The energy density, for instance,
has been found to rise rapidly at $T_c$ and approach the high
temperature ideal gas limit from below. 
Such calculations are performed on finite Euclidean space-time lattices  
of size $N_\sigma^3 \times N_\tau$ with non-zero lattice spacing $a$, which
is controlled through the dimensionless coupling $\beta = 2N/g^2$. This
fixes also the thermodynamic parameters, temperature $T=1/N_\tau a$ and
volume $V=(N_\sigma a)^3$. In general such calculations suffer from two
different types of finite size errors. An infra-red cut-off is introduced
through the finite spatial size $N_\sigma$ and an ultra-violet cut-off, at
fixed temperature, is introduced through the temporal extent $N_\tau$. 
While $N_\sigma$ should be large in order to reach the thermodynamic limit
($V\rightarrow \infty$), $N_\tau$ should be large in order to eliminate the
influence of the regularization (finite lattice cut-off) on physical
observables in the continuum limit. In the following we mainly will be 
concerned with the latter aspect.

As a consequence of the $O(a^2)$ discretization errors in the standard
Wilson formulation of $SU(N)$ gauge theories thermodynamic quantities
calculated on the lattice receive $O((aT)^2)$ corrections.
For an ideal gluon gas these corrections are known to be as large as 50\% on
lattices with $N_\tau = 4$, ie. $aT\equiv N_\tau^{-1}$.  Asymptotically, in 
the limit $g^2 \rightarrow 0$, the cut-off dependence of energy density and
pressure are given by \cite{Eng95,Bei95},
\beqn
{\epsilon \over T^4} = 3{p \over T^4} =
(N^2-1) {\pi^2 \over 15}\biggl[1+
+ {30 \over 63} \cdot {\pi^2 \over N_\tau^2} +
 {1 \over 3} \cdot {\pi^4 \over N_\tau^4} +
O\biggl ({1 \over N_\tau^6} \biggr ) \biggr ]~.
\label{Ntdependence}
\eqn

Cut-off effects are
particularly large in observables, which are sensitive to the ultra-violet
behaviour of the theory. This is the case for bulk thermodynamic
quantities in the high temperature limit.
As will be discussed in the following these finite cut-off effects are, in fact, 
clearly visible in the calculations of
the pressure and energy density performed on lattices with varying temporal
extent, although in the temperature regime up to a few times $T_c$ they turn
out to be about a factor 2 smaller in magnitude than calculated analytically
in the $T \rightarrow \infty$ limit \cite{Boyd95}. 
Only recently these calculations could be extended to lattices
with sufficiently large temporal extent
($N_\tau =6$ and 8) that would allow
an extrapolation of lattice results for bulk
thermodynamic quantities to the continuum limit \cite{Eng95,Boyd95}.

On the other hand,
cut-off effects are of less importance for the analysis of long-distance 
properties of the theory, such as the string tension and the deconfinement
transition temperature itself. In this case it is more important
to reach the large volume (thermodynamic) limit.
The critical temperature is calculated at finite lattice
cut-off, ie. on lattices with fixed temporal extend $N_\tau$, by
determining the corresponding critical couplings $\beta_c (N_\tau)$
\cite{Boyd95,Fin93,Iwa92,Christ}. A
calculation of the string-tension at this value of the coupling on a large
zero temperature lattice is then used to set the scale,
\beqn
\biggl( {T_c \over \sqrt{\sigma} }\biggr)_{|a\equiv 0} =
\biggl({T_c \over \sqrt{\sigma}  }\biggr)_{|aT=1/N_\tau} +
c_2 (\sigma a^2) ~~.
\label{tcsfit}
\eqn

\begin{figure}[htb]
\centerline{
   \epsfig{bbllx=115,bblly=280,bburx=480,bbury=635,
       file=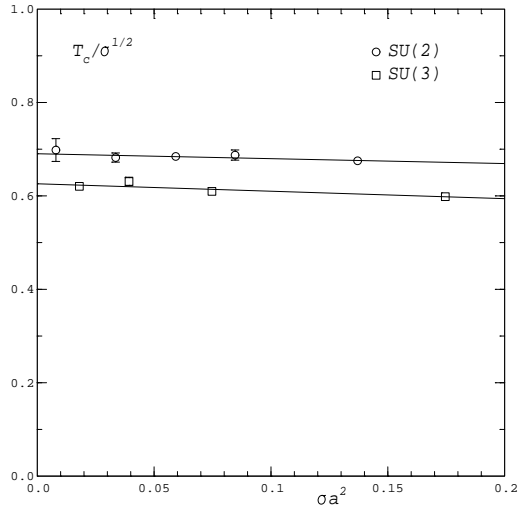,width=70mm}}
\caption{\baselineskip 10pt
The critical temperature in units of the square root of the
string tension for the $SU(2)$ and $SU(3)$ gauge theories versus the lattice
cut-off squared.}
\label{fig:tc}
\end{figure}

A collection of results for this ratio is shown for the $SU(2)$ and $SU(3)$ 
gauge theories in Figure~\ref{fig:tc}. 
As can be seen the $O(a^2)$ corrections are quite
small. An extrapolation to the continuum limit yields
\beqn
{T_c \over \sqrt{\sigma}} = \cases{0.69 \pm 0.01 &,\quad SU(2) \cr
0.629 \pm 0.003(+0.004) &,\quad SU(3) }~,
\label{Tcratio}
\eqn
where the number in brackets, given in the case of $SU(3)$, indicates the
systematic error still present in the data due to the missing extrapolation
to the thermodynamic limit on the larger lattices.

\begin{figure}[htb]
\begin{center}
  \epsfig{bbllx=80,bblly=180,bburx=515,bbury=610,
       file=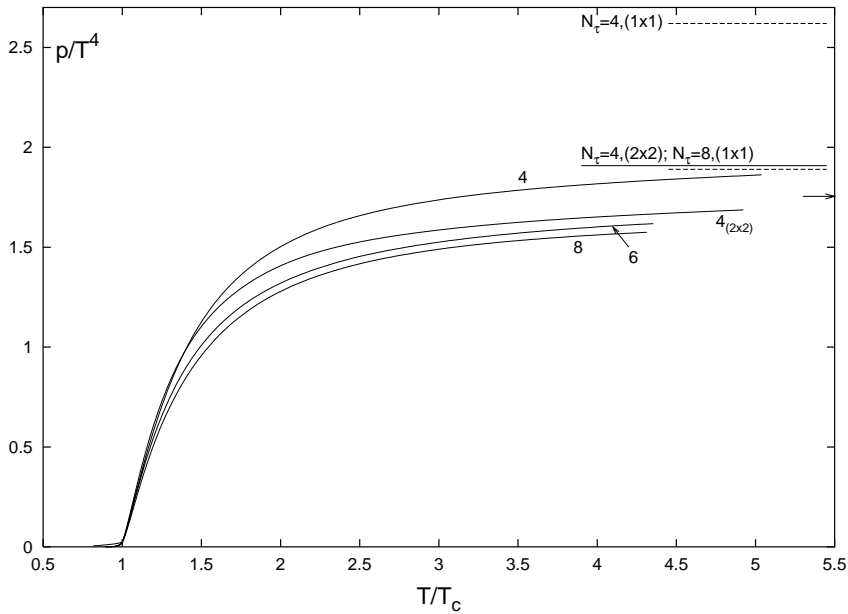, height=70mm, angle=-90}
\end{center}
\vskip 2pt 
\caption{\baselineskip 10pt
The pressure versus $T/T_c$ for $N_\tau = 4$, 6 and 8
obtained from an integration
of the action density for the standard one-plaquette Wilson action. For 
$N_\tau = 4$ we also show the result obtained with an improved action
($4_{(2\times 2)})$. }
\label{fig:pressure}
\end{figure}

Let us now turn to a discussion of bulk thermodynamic observables.
The pressure can be obtained from an integration of the difference of
action densities at zero
($S_0$) and finite ($S_T$) temperature \cite{Eng90},
\beqn
{p\over T^4}\Big\vert_{\beta_0}^{\beta}
=~N_\tau^4\int_{\beta_0}^{\beta}
{\rm d}\beta'  (S_0-S_T) ~,
\label{freelat}
\eqn
where the action densities themselves may be chosen appropriately in order
to reduce cut-off effects. The simplest choice corresponds to the standard
Wilson action, which leads to $O((aT)^2)$ errors in thermodynamic
observables. It is given in terms of plaquette expectation values,
\beqn
S^{(1\times 1)}_{0(T)}= 6 \langle 1-{1\over 3}{\rm Tr} W^{(1\times 1)} 
\rangle_{0(T)}~~,
\label{wilsonaction}
\eqn
with $W^{(1\times 1)}$ denoting the product of the four link-variables
around an elementary plaquette in the lattice.
Various improved actions have been suggested, which reduce the cut-off
effects to $O(((aT)^4)$. One possibility is to add to the single plaquette
term in the Wilson action a 
contribution of $(2\times 2)$-loops, $W^{(2\times 2)}$, 
(for further details see \cite{Bei95}),
\beqn
S^{((2\times 2)}_{0(T)}= 6\biggl[ {4\over 3} 
\langle 1-{1\over 3}{\rm Tr}W^{(1\times 1)} \rangle_{0(T)} -{1\over 48}
\langle 1-{1\over 3}{\rm Tr}W^{(2\times 2)} \rangle_{0(T)} \biggr]~~.
\label{improvedaction}
\eqn
Using either of these actions in the simulation one can perform the 
numerical integration defined in Eq.~\ref{freelat}. This yields
the pressure difference between two temperatures,
corresponding to the two couplings $\beta_0$ and $\beta$.
The lower temperature, corresponding to $\beta_0$, can be chosen small enough
so that the pressure can be approximated by zero at this point. As the low 
temperature
phase in a $SU(N)$ gauge theory is a glueball gas, ie. a system with rather heavy
constituents, the pressure drops exponentially for
$T< T_c$. The integration therefore can be stopped rather close to $T_c$.
Results obtained for the pressure with the standard Wilson action 
on lattices with temporal extent $N_\tau =4$, 6 and 8 as well as with an
improved action on a $N_\tau =4$ lattice are shown in Figure~\ref{fig:pressure}. 
One clearly sees the expected cut-off dependence. As discussed above, it
qualitatively reflects the $N_\tau$-dependence of the free gluon gas, which
also is shown by horizontal lines in this figure. We note that the
improved action calculation, performed on a lattice with only four sides in
the temporal direction indeed leads to a smaller cut-off dependence than the
corresponding standard Wilson action. 

\begin{figure}[htb]
\begin{center}
  \epsfig{bbllx=80,bblly=180,bburx=515,bbury=610,
       file=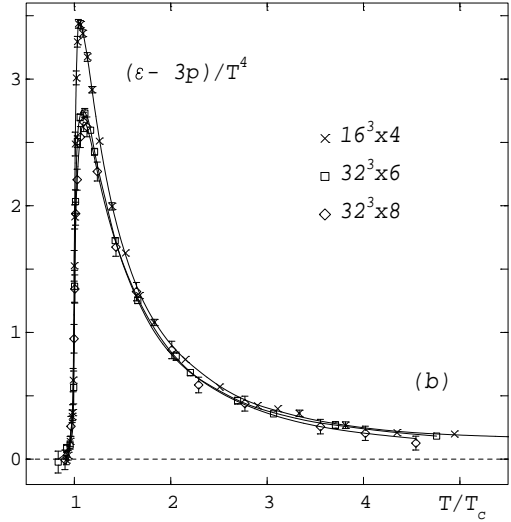, height=85mm}
\end{center}
\caption{\baselineskip 10pt
The difference $(\epsilon -3p)/T^4$ obtained from calculations with the
standard Wilson action on lattices with temporal extent
$N_\tau = 4$, 6 and 8.}
\label{fig:e3p}
\end{figure}

Making use of basic thermodynamic relations we can then evaluate the
energy density in the thermodynamic limit from the trace anomaly,
\beqn
{\epsilon - 3p \over T^4}  =  T {\partial \over \partial T} (p/T^4) =
 -6N_\tau^4 \biggl( a{\partial g^{-2} \over \partial a} \biggr)
\biggl( S_0-S_T \biggr)~~,
\label{epsilonlat}
\eqn
where the derivative $a\partial g^{-2} /\partial a$ is obtained from an
explicit parameterization of the dependence of the cut-off, $a$, on the
bare coupling $g^2$ \cite{Boyd95}. This is shown in Figure~\ref{fig:e3p}.

The much slower rise of the pressure relative to the energy density
results in a large peak in the trace anomaly, $(\epsilon -3p)/T^4$,
at $T_{\rm peak}\simeq 1.1 T_c$. The trace anomaly defines,
the difference between the gluon condensate at zero and finite temperature
$\epsilon -3p = G^2(0) -G^2(T)$. From Figure~\ref{fig:e3p} 
we find at $T_{\rm peak}$,
\beqn
(\epsilon -3p )_{\rm peak} = (0.57\pm 0.02) \sigma^2 \simeq 2.3~{\rm
GeV/fm}^3 ,
\label{condensate}
\eqn
where we have used the result given in
Eq.~\ref{Tcratio} for the critical temperature in units of the string tension 
and $\sqrt{\sigma} \simeq 420$~MeV in
order to set the physical scale. This result is, in fact, compatible with
conventional values
for the zero temperature gluon condensate, $G^2(0) \simeq 2~$GeV/fm$^3$
\cite{Leu92}. It, therefore, 
suggests that $G^2(T) \simeq 0$ at $T_{\rm peak}$.

\vskip 20pt
\noindent
{\large \bf II.3 \hskip 0.3truecm Extrapolation to the continuum limit}
\vskip 5pt

\begin{figure}[htb]
\begin{center}
  \epsfig{bbllx=80,bblly=90,bburx=515,bbury=690,
       file=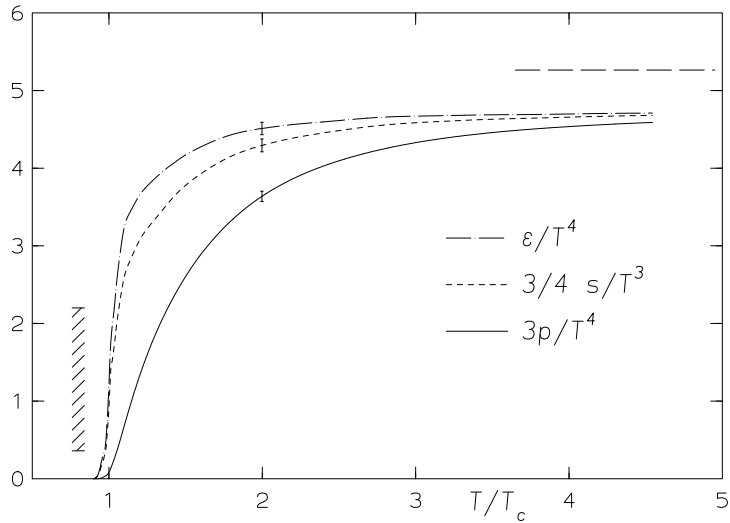, height=100mm, angle=-90}
\end{center}
\caption{\baselineskip 10pt
Extrapolation to the continuum limit for the energy density,
entropy density and pressure versus $T/T_c$. The dashed horizontal line
shows the ideal gas limit. The hatched vertical band indicates the size of the
discontinuity in $\epsilon/T^4$ (latent heat) at $T_c$ [10]. Typical error
bars are shown for all curves.}
\label{fig:continuum}
\end{figure}

Based on the analysis of the pressure and energy density on various size
lattices one can now attempt to extrapolate these quantities to the 
continuum limit. At present this can only be done for the Wilson action 
where results do exist on lattices with temporal extent up to $N_\tau = 8$.
Using an ansatz motivated by Eq.~\ref{Ntdependence},
\beqn
\biggl({p \over T^4}\biggr)_{N_\tau} = \biggl({p \over T^4}\biggr)_{\infty}
+ {{\rm const.} \over N_\tau^2} ~~,
\label{cfit}
\eqn
we extrapolate the $N_\tau =6$ and 8 data, respectively \cite{Boyd95}.
The resulting continuum predictions for
the pressure, energy density and entropy density
are shown in Figure~\ref{fig:continuum}.
We generally find that the difference between the extrapolated values and
the results for $N_\tau=8$ is less than 4\%, which should be compared with
the corresponding result for the free gas, where the difference is still
about 8\%. This suggests that relative to the ideal gas case more low
momentum modes, which are less sensitive to finite cut-off effects, contribute
to thermodynamic quantities.

\begin{figure}[htb]
\begin{center}
  \epsfig{bbllx=90,bblly=180,bburx=505,bbury=600,
       file=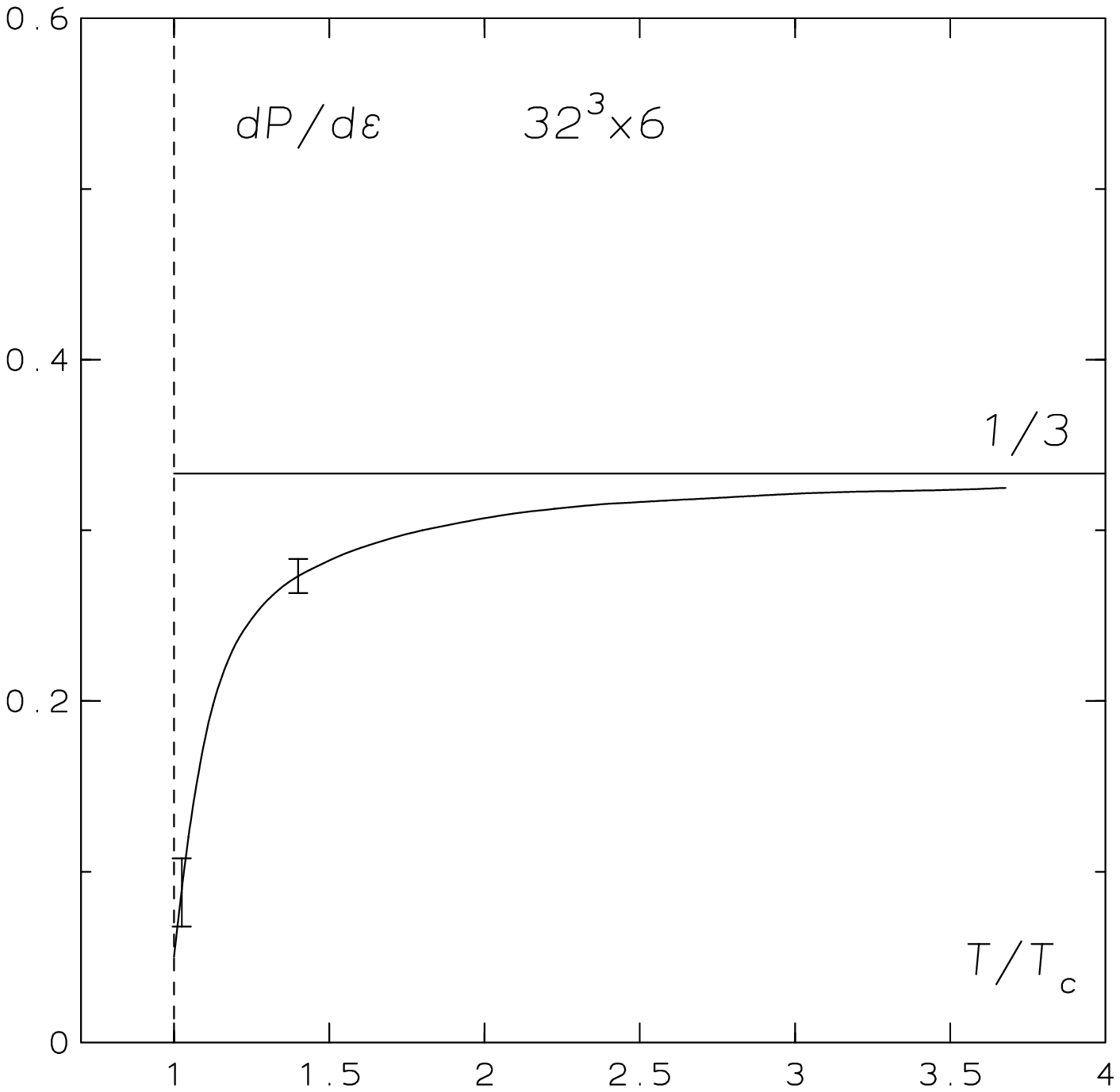, height=70mm}
\end{center}
\caption{\baselineskip 10pt
Velocity of sound for the $SU(3)$ gauge theory calculated on a lattice
with temporal extent $N_\tau =6$. We do not observe any significant cut-off
dependence for this derivative.}
\label{fig:sound}
\end{figure}

Knowing the energy density and pressure as a function of temperature we also
can evaluate other thermodynamic quantities. A quantity of particular
interest for the description of the hydrodynamic expansion of a quark-gluon plasma
created in a heavy ion collision is the velocity of sound,
\beqn
v^2 = {{\rm d} p \over {\rm d}\epsilon}~~.
\label{sound}
\eqn
Already from Figure~\ref{fig:continuum} it is clear that $v^2$ will be small
close to $T_c$ because the pressure changes only little there, while the energy
density rises rapidly. This, indeed, is seen in Figure~\ref{fig:sound}. We
also note that the velocity of sound rapidly approaches the ideal gas limit,
$v^2_{\rm ideal} = 1/3$.

The above discussion of energy density and pressure in a $SU(N)$ gauge
theory has shown, that both quantities approach the ideal gas limit from
below. We note that
the energy density rapidly rises to about 85\% of the ideal gas value at
$2T_c$ and then shows a rather slow increase, which is consistent with a
logarithmic increase as one would expect if the coupling $g(T)$ changes only
logarithmically with temperature. 

We have not discussed here the status of the equation of state for QCD with
light quarks. A detailed investigation for two flavour QCD has been
performed recently on lattices with temporal extent $N_\tau=4$ \cite{Blu95}.
However, also here an analysis on larger lattices is still needed in order to
control the obviously present large discretization errors.

\vskip 20pt
\noindent
{\large \bf III \hskip 0.3truecm QCD with Light Quarks}
\vskip 5pt

\noindent
{\large \bf III.1 \hskip 0.3truecm The chiral phase transition}
\vskip 5pt

\noindent
A well established result from lattice calculations is the strong
dependence of the order of the phase transition on the QCD symmetries
related to the colour as well as the flavour degrees of freedom. In the
absence of quarks, ie. in pure $SU(N)$ gauge theories, the phase
transition is second order for $N=2$ and first order for $N=3$.
In the case of QCD with $n_f$ light quark flavours
the transition is first order for $n_f \ge 3$ and does seem to be
continuous for $n_f=2$. Lattice calculations also show a strong
dependence of the transition temperature on the number of partonic
degrees of freedom. The transition temperature, expressed for instance
in units of
the square root of the string tension ($T_c/\sqrt{\sigma}$), is
substantially smaller in QCD with light quarks than in a pure
gauge theory. In the case of QCD with two light quarks one
finds $T_c(n_f=2) \simeq 150$MeV, while in the purely gluonic
theory, $n_f \equiv 0$, $T_c$ is found to be
substantially higher, $T_c(n_f=0) \simeq 260$MeV. As discussed in the
previous section this may be understood in terms of a percolation
threshold:
In the absence of quarks there are no light hadrons, the lightest
states are glueballs with a mass of $O(1~{\rm GeV})$. Therefore, a
rather high temperature is needed to excite enough of these
heavy glueball states in order to reach a critical density where
hadrons start overlapping.

With increasing temperature non-perturbative properties of the QCD
vacuum, like confinement and chiral symmetry breaking, get modified
through the thermal heat bath and
eventually disappear above $T_c$.
Non-vanishing condensates  play an important role in understanding
these non-perturbative features of QCD. They vary  with temperature
in the vicinity of the QCD phase transition.
The restoration of chiral symmetry at $T_c$
will, for instance, be reflected in a characteristic temperature dependence
of the chiral condensate, $\langle \bar\psi \psi \rangle \sim
|T-T_c|^{\beta}$, which also
should have consequences for the properties of hadrons close to $T_c$.
Effective theories, deduced from QCD in order to describe low energy
properties of hadrons, establish a close link between hadronic
properties and the non-perturbative structure of the QCD vacuum.
This is, for instance,
used in the operator product expansion (OPE) for correlation functions
of hadronic currents \cite{Shi93}, which allows to relate hadron masses to
condensates of the quark and gluon fields. The temperature and density
dependence of the latter has been discussed within the context of chiral
perturbation theory \cite{Leu92}.

We will discuss here some of the recent results of lattice calculations
on the phase transition in two-flavour QCD as well as
an analysis of hadronic properties close to the QCD phase
transition. Our discussion is limited to the staggered formulation of
lattice regularized QCD. First attempts to study the chiral 
symmetry restoration also with Wilson fermions have been reviewed in
\cite{Kan95}. In the next section we briefly
discuss the critical behaviour in two-flavour QCD.
Section III.3 is devoted to a
discussion of lattice calculations of hadron masses and decay constants
at finite temperature within the quenched approximation.
\vskip 10pt
\noindent
{\large \bf III.2 \hskip 0.3truecm The chiral transition in two-flavour QCD \hfil}
\vskip 5pt

\noindent
The nature of the chiral phase transition does seem to depend
strongly on the number of light quark flavours, $n_f$.
The special role of two-flavour QCD has been noticed
already some time ago by Pisarski and Wilczek \cite{Pis84}. They
have suggested that the dynamics of the chiral phase transition is
controlled by an effective, 3-$d$ scalar
Lagrangian constructed in terms of the chiral order parameter field,
$\sigma \sim \bar\psi \psi$. This
effective theory has a global $O(4)$ symmetry ($SU(2)\times
SU(2)$), which
is spontaneously broken to $SU(2)$. It therefore is expected that in
the case of
a continuous transition the critical behaviour is governed by critical
exponents of a 3-$d$, $O(4)$ symmetric spin model \cite{Pis84,Wil92}.
A remaining question in this scenario is
to what extent the axial $U(1)$ influences the
critical behaviour in the vicinity of $T_c$ \cite{Wil92,Shu94}.
An even more fundamental problem is whether an effective scalar
theory can at all describe the critical behaviour of a theory,
in which the scalar fields are only constructed as
fermionic bilinears. It recently has been argued by Koci\'c and
Kogut that this may not be the case and that two-flavour QCD may
instead show mean field behaviour in the vicinity of $T_c$ \cite{Koc94}.
In lattice calculations an additional problem arises: None of the
fermion
Lagrangians used for numerical calculations has the complete chiral
symmetry of the continuum Lagrangian. In the staggered fermion
formulation, for instance, one has a chiral $U(1)$ rather than the full
$SU(2)$ symmetry, which only is restored in the continuum limit. This
too may influence the critical behaviour in numerical studies.
A detailed quantitative investigation of the critical behaviour thus is
needed in different lattice formulations. Here we will restrict our
discussion to the case of staggered fermions.

\begin{figure}[htb]
\begin{center}
    \leavevmode
    \epsfig{bbllx=160,bblly=320,bburx=500,bbury=520,
        file=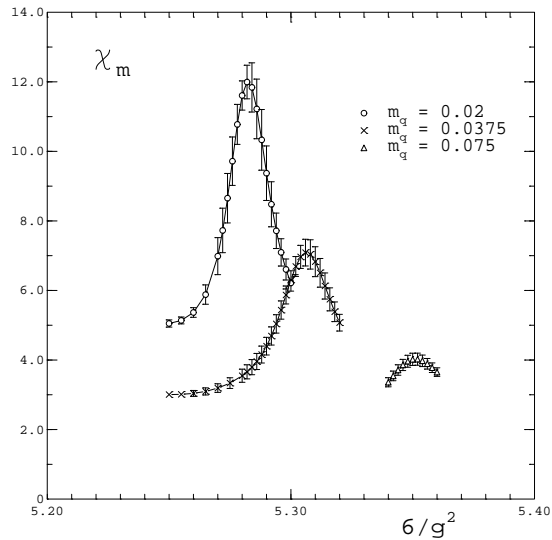,height=40mm,angle=-90}
\end{center}
\vskip -0.8truecm
\caption{The chiral susceptibility in two-flavour QCD versus $6/g^2$ for
three values of the quark mass on a $8^3\times 4$ lattice.}
\label{fig:suscept}
\end{figure}

The critical behaviour of QCD with light quarks is controlled by two
dimensionless parameters, the reduced temperature $t=(T-T_c)/T_c$ and
the scaled quark mass, $h=m_q/T$. The latter corresponds to an external
magnetic field in spin models. In the case of QCD with dynamical
quarks it is the quark mass rather than the spatial lattice volume,
which is the most stringent limitation for large correlation lengths in the
vicinity of the critical point. Rather than performing a finite size
scaling analysis, as it has been done in the case of the deconfinement
transition in $SU(N)$ gauge theories, it thus is more appropriate to
consider the
scaling behaviour in terms of the quark mass. Finite volume effects may
in a first approximation be ignored.

In the vicinity of the critical point the behaviour of bulk
thermodynamic quantities is related to thermal ($y_t$) and
magnetic $(y_h)$ critical exponents, which characterize the
scaling behaviour of the singular part of the free energy
density,
\begin{equation}
f_s(t,h)\equiv - {T \over V} \ln Z_s =b^{-1}f(b^{y_t}t, b^{y_h} h)~~.
\label{free}
\end{equation}
Here $b$ is an arbitrary scale factor. Various scaling relations for
thermodynamic quantities can be derived from Eq.~\ref{free}
\cite{Kar94,KarL94}. For instance, one finds for the chiral order
parameter, $\langle \bar\psi \psi \rangle$, and its derivative with
respect to the quark mass (the chiral susceptibility $\chi_m$)
\begin{eqnarray}
\langle \bar\psi \psi \rangle (t,h) &= & h^{1/\delta} F(z)
\nonumber \\
\chi_m (t,h) &= & {1\over \delta} h^{1/\delta - 1} \biggl[ F(z) -
{z \over
\beta} F'(z) \biggr]~~,
\label{scale}
\end{eqnarray}
with scaling functions $F$ and $F'$ that only depend on a
specific combination of the reduced temperature  and scaled
quark mass, $z=th^{-1/ \beta\delta}$. The critical
exponents $\beta$ and $\delta$ are
given in terms of $y_t$ and $y_h$ as  $\beta = (1-y_h)/y_t$ and
$\delta = y_h/(1-y_h)$. A direct consequence of Eq.~\ref{scale}
is, for
instance, that $\chi_m$ has a peak located at some fixed value of
$z_c$. This defines a pseudo-critical point, $t_c(h)\equiv z_c
h^{1/ \beta\delta}$, at which the peak height of $\chi_m$ increases
with decreasing quark mass according to Eq.~\ref{scale}.
Fig.~\ref{fig:suscept} shows the scaling behaviour of $\chi_m$
for three values of the quark mass \cite{KarL94}.
The rapid rise of $\chi_m$ with decreasing quark mass can be used to
determine the critical exponent $\delta$.

\begin{figure}[htb]
\begin{minipage}[t]{80mm}
  \begin{center}
    \leavevmode
    \epsfig{bbllx=160,bblly=320,bburx=500,bbury=520,
        file=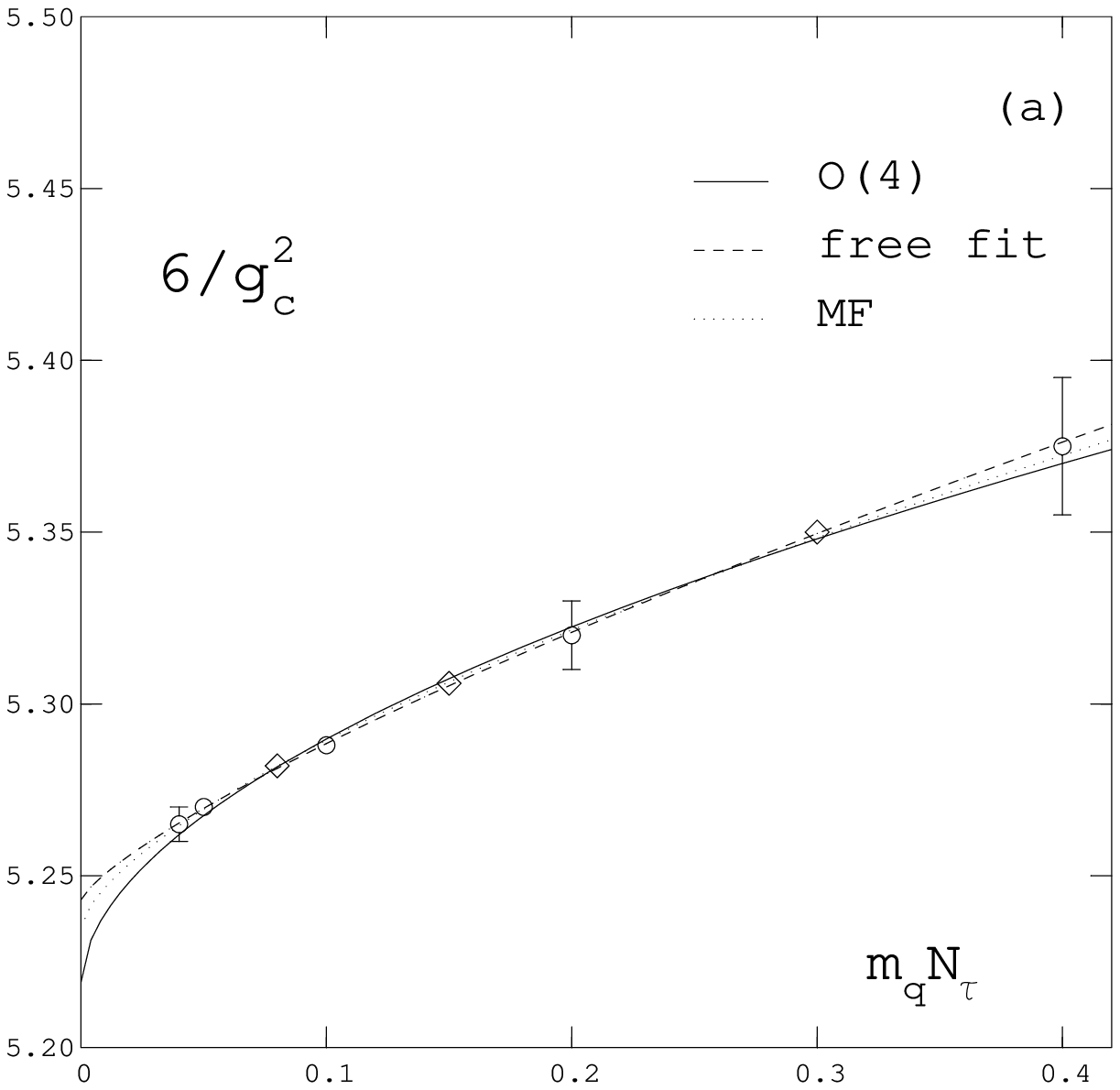,height=40mm,angle=-90}
   \end{center}
\end{minipage}
%
%
\begin{minipage}[t]{75mm}
  \begin{center}
    \leavevmode
    \epsfig{bbllx=160,bblly=320,bburx=500,bbury=520,
        file=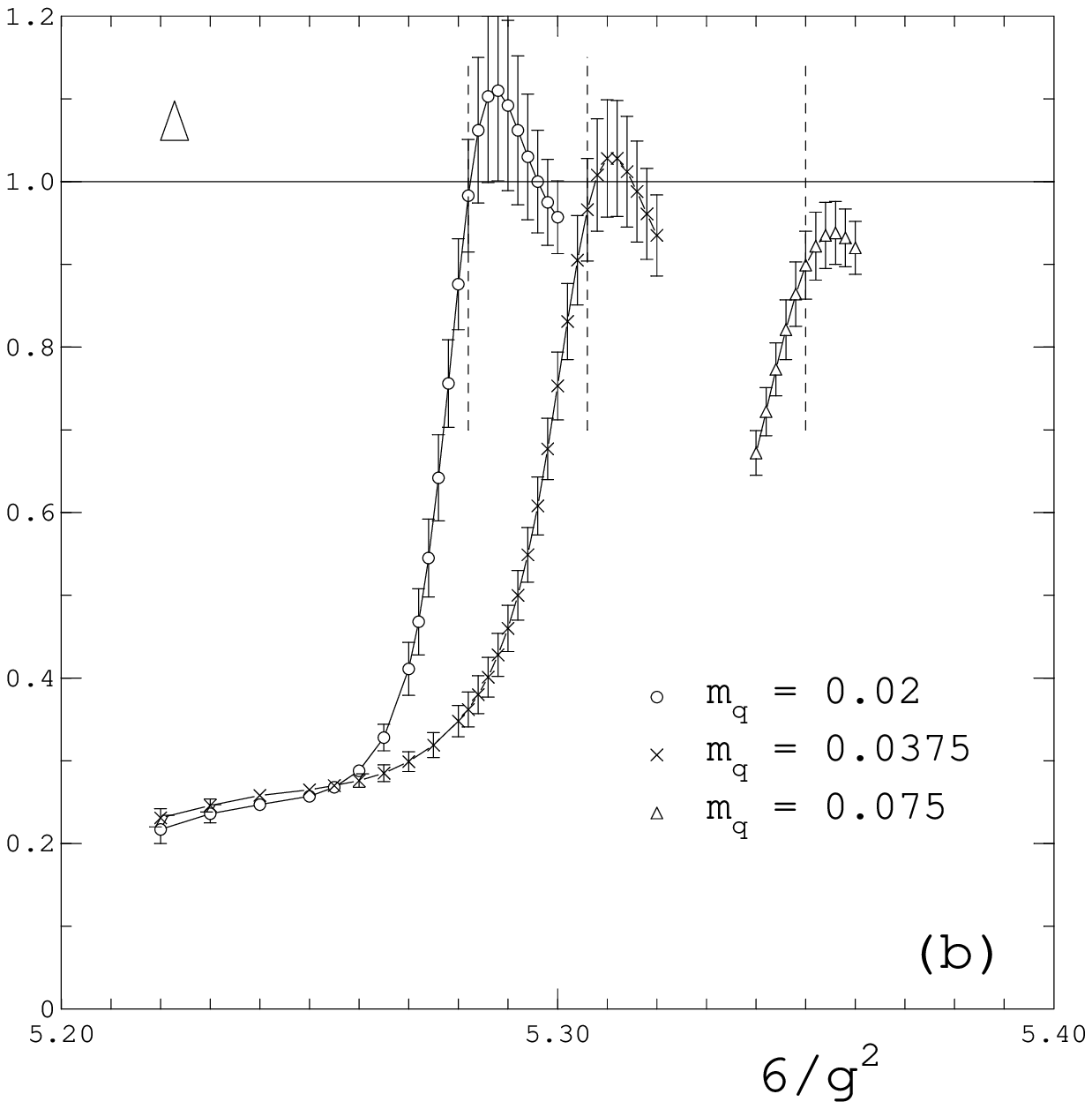,height=40mm,angle=-90}
   \end{center}
\end{minipage}
\caption{The pseudo-critical couplings (a)
of the chiral transition in two-flavour QCD on lattices with
temporal extent $N_\tau=4$ and the chiral cumulant (b).
Shown are results from calculations with
different values of the quark mass on a $8^3\times 4$ lattice.}
\label{fig:coupling}
\end{figure}

Another useful observable for the determination of the critical exponent
$\delta$ is the chiral cumulant,
\begin{equation}
\Delta (z) = {h \chi_m \over \langle \bar\chi \chi \rangle}~~.
\label{cumulant}
\end{equation}
From Eq.~\ref{scale} it follows that $\Delta$ is directly a scaling
function, ie. it only depends on $z$. Moreover, it is uniquely
determined at $z=0$, where it takes on the value $1/\delta$ even at
non-vanishing values of the quark mass. The chiral cumulant thus allows a
direct calculation of this critical exponent.
In Fig.~\ref{fig:coupling}a we show
a collection of the pseudo-critical couplings obtained from Monte
Carlo calculations with various values of the quark mass and on
various different lattice sizes. The curves show various fits for the quark
mass dependence of these pseudo-critical couplings. Apparently it is difficult
to distinguish $O(4)$ exponents from mean field exponents in this way. We
note, however, that the location of the zero quark mass critical
coupling is quite well determined. The cumulant $\Delta$ is shown in
Fig.~\ref{fig:coupling}b. For the two smaller quark masses this quantity
has been evaluated in the region of the estimated zero quark mass critical
coupling, $\beta_c(m_q=0) \simeq 5.243(10)$. In this interval the cumulant
takes on values between 0.21 and 0.26, which is compatible with the $O(4)$
exponent $1/\delta = 0.205(45)$.

Further scaling relations can be obtained by considering the quark
mass dependence of the thermal susceptibility, $\chi_t \sim \partial
\langle \bar\psi \psi \rangle / \partial t$, and the specific heat. A
combined analysis of these allows a determination of the critical
exponents $y_t$ and $y_h$. Results from a first analysis of this
kind \cite{KarL94} are summarized in Table~\ref{tab:exp} and compared with
the known exponents for 3-$d$, $O(4)$ symmetric spin models \cite{Kan94} and
mean field behaviour.

\begin{table}[hbt]
\newlength{\digitwidth} \settowidth{\digitwidth}{\rm 0}
\catcode`?=\active \def?{\kern\digitwidth}
\caption{Thermal ($y_t$) and magnetic ($y_h$) exponents from
simulations of
two-flavour QCD on an $8^3\times 4$ lattice (LGT).
Also given are mean field exponents (MF)
and those for a 3-$d$, $O(4)$ symmetric spin model.}
\label{tab:exp}
\begin{tabular*}{\textwidth}{@{}l@{\extracolsep{\fill}}rrrr}
\hline
                 & LGT & $O(4)$ & MF  \\
\hline
$y_t$            & $0.69(7)$  & $0.446(5)$ & $0.5$  \\
$y_h$            & $0.83(3)$  & $0.830(6)$ & $0.75$  \\
\hline
\end{tabular*}
\end{table}
As can be seen from the Table the exponent $y_h$ extracted from
lattice calculations is in very good agreement with the expected $O(4)$
model behaviour. It corresponds to a value of 0.205(45) for the exponent
$1/\delta$, which is significantly below the value 1/3 expected for a
mean field critical exponent.
The exponent $y_t$, however, comes out too large in the present calculation.
It may be expected that the determination of $y_t$ is more sensitive to finite
volume effects than that of $y_h$. In the limit of vanishing quark
masses it is $y_t$ that controls the finite volume scaling behaviour of
thermodynamic quantities.

Clearly the analysis of critical exponents on a $8^3 \times 4$ lattice
can only be considered as a first exploratory
step towards a detailed investigation of critical exponents in
two-flavour QCD. In the near future calculations on larger lattices
with smaller quark masses will certainly lead
to much better determinations of critical exponents. This should allow
to distinguish between mean field and $O(4)$ critical behaviour.

The chiral susceptibility,
$\chi_m$, has a very narrow peak at the pseudo-critical point.
This indicates that only in this region contributions from the singular
part of the free energy, which is responsible for the occurrence of a phase
transition, is dominant. This behaviour is consistent with the
temperature dependence of $\langle \bar\psi \psi \rangle$,
which leads to deviations from the zero temperature condensate only close
to $T_c$.
We thus expect that also hadronic properties will show significant
modifications  only in a narrow temperature
regime close to $T_c(m_q)\equiv T_{\rm peak}$,
\beqn
\biggl| {T_{\rm peak/2}-T_{\rm peak} \over T_{\rm peak}}
\biggr| \lsim 0.1~~.
\label{crit}
\eqn

\vskip 10pt
\noindent
{\large \bf III.3 \hskip 0.3truecm  Hadronic properties close to $T_c$ \hfil}
\vskip 5pt

\noindent
Information about hadron masses and decay constants can be extracted in
lattice simulations from the long-distance behaviour of correlation
functions of hadron operators
\beqn
G_H(x) = \langle H(x) H^+ (0) \rangle \rightarrow e^{-m_H |x|}
\quad,\quad x\equiv (\tau, \vec x)~~,
\label{corr}
\eqn
where $H(x)$ denotes an operator with the appropriate quantum numbers of
the hadronic state under consideration, for instance $H(x) = \bar\psi_u(x)
\gamma_5 \psi_d(x)$ for the pion. At zero temperature one studies
the behaviour of the correlation function at large Euclidean times $\tau$.
From the exponential decay of $G_H$ one deduces the hadron mass of the
lightest state in this channel, whereas the amplitude is related to the
decay constant of this hadronic state. At finite temperature the Euclidean
extent is limited, $0\le \tau \le 1/T$, and one thus studies the behaviour
of $G_H$ for large spatial separations, $|\vec x | \rightarrow \infty$.
This yields information about the finite temperature screening masses
which are related to pole masses as long as there is a bound state
in the quantum number channel under consideration.

Finite temperature screening masses have been studied in lattice
simulations of QCD for quite some time \cite{DeT87}. A very drastic qualitative
change in the screening masses is seen when one crosses the QCD
transition temperature. Parity partners become degenerate above $T_c$,
the pseudo-scalar mass becomes massive and approaches the free
quark/anti-quark value, $m_{\rm meson} = 2\pi T$, at large temperatures.
These features do not seem to depend much on the number of quark
flavours. In particular, they also have been found in quenched QCD
simulations. It thus seems to be meaningful to first study the temperature
dependence
of hadronic properties in the quenched approximation where results on
large lattices with high accuracy can be obtained. In the following
we will describe the results of such an investigation performed on a
rather large lattice ($32^3\times 8$) close to the phase transition
temperature \cite{Boy94}.
\vskip 10pt
\noindent
{\large \bf III.3.1 \hskip 0.3truecm  The GMOR relation and the pion decay constant
\hfil}
\vskip 5pt

\noindent
The GMOR relation gives the chiral condensate in terms of the pion mass
and pion decay constant,
\beqn
f_\pi^2 m_\pi^2 = m_q \langle \bar\psi \psi \rangle_{m_q=0}~~.
\label{fp}
\eqn
Below $T_c$ the pion is a Goldstone particle, its
mass squared depends linearly on the quark mass,
\beqn
m_\pi^2 = a_\pi m_q ~~.
\label{mp}
\eqn
A calculation of the chiral condensate and the pion mass at different
values of the quark mass allows the determination of the pion slope,
$a_\pi$, and the zero quark mass limit of the condensate,
$\langle \bar\psi \psi \rangle_{m_q=0}$.

The pion decay constant $f_\pi$ can be determined directly from the
relevant matrix element
\beqn
\sqrt{2} f_\pi m_\pi^2 = \langle 0| \bar\psi_u \gamma_5 \psi_d |\pi^+
\rangle ~~.
\label{fpmp}
\eqn

\begin{figure}[htb]
\begin{minipage}[t]{80mm}
  \begin{center}
\vskip 0.01truecm
    \leavevmode
    \epsfig{bbllx=100,bblly=245,bburx=450,bbury=600,
        file=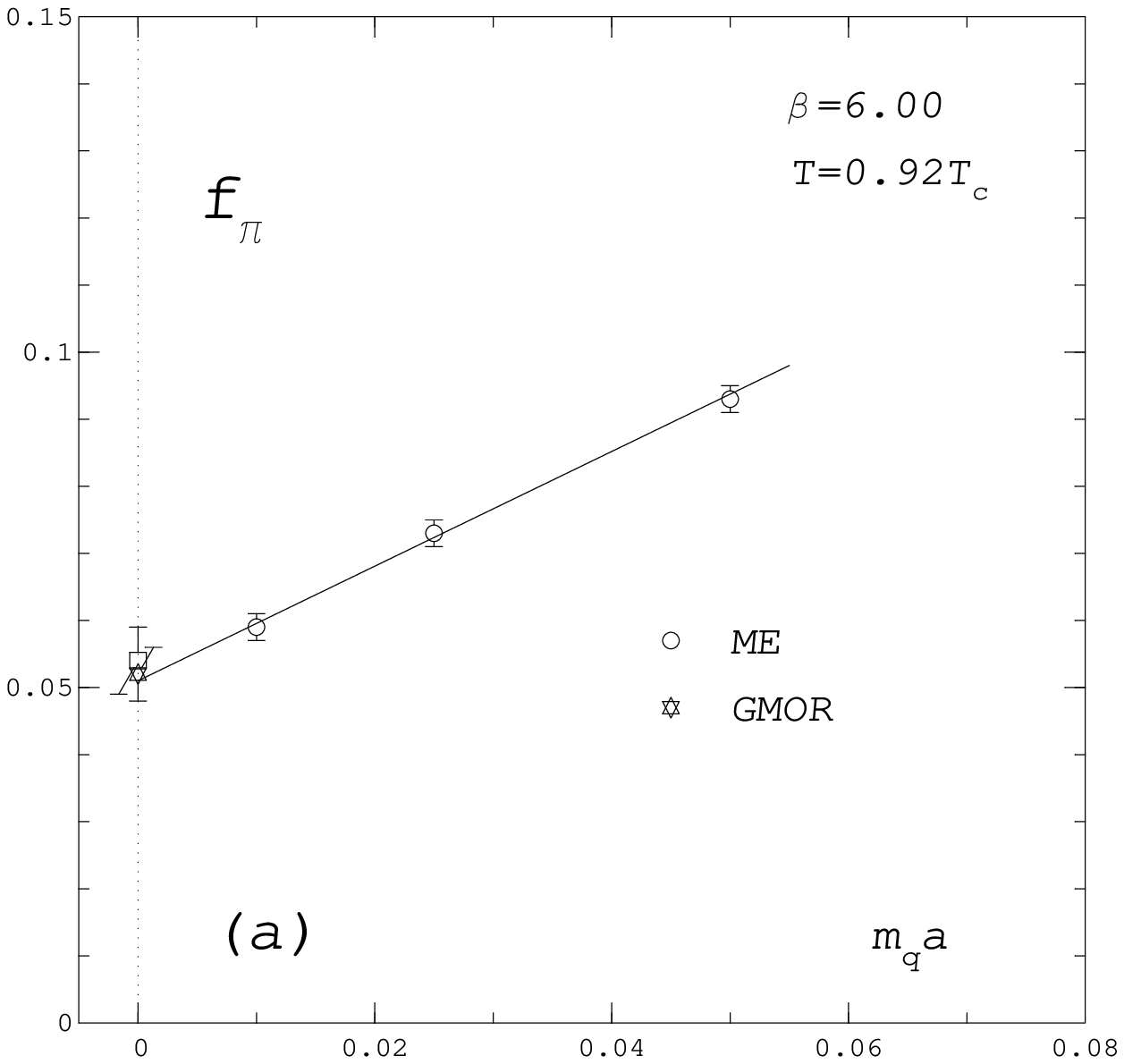,height=70mm}
   \end{center}
\end{minipage}
%
%
\begin{minipage}[t]{75mm}
  \begin{center}
\vskip -0.6truecm
    \leavevmode
    \epsfig{bbllx=100,bblly=245,bburx=450,bbury=600,
        file=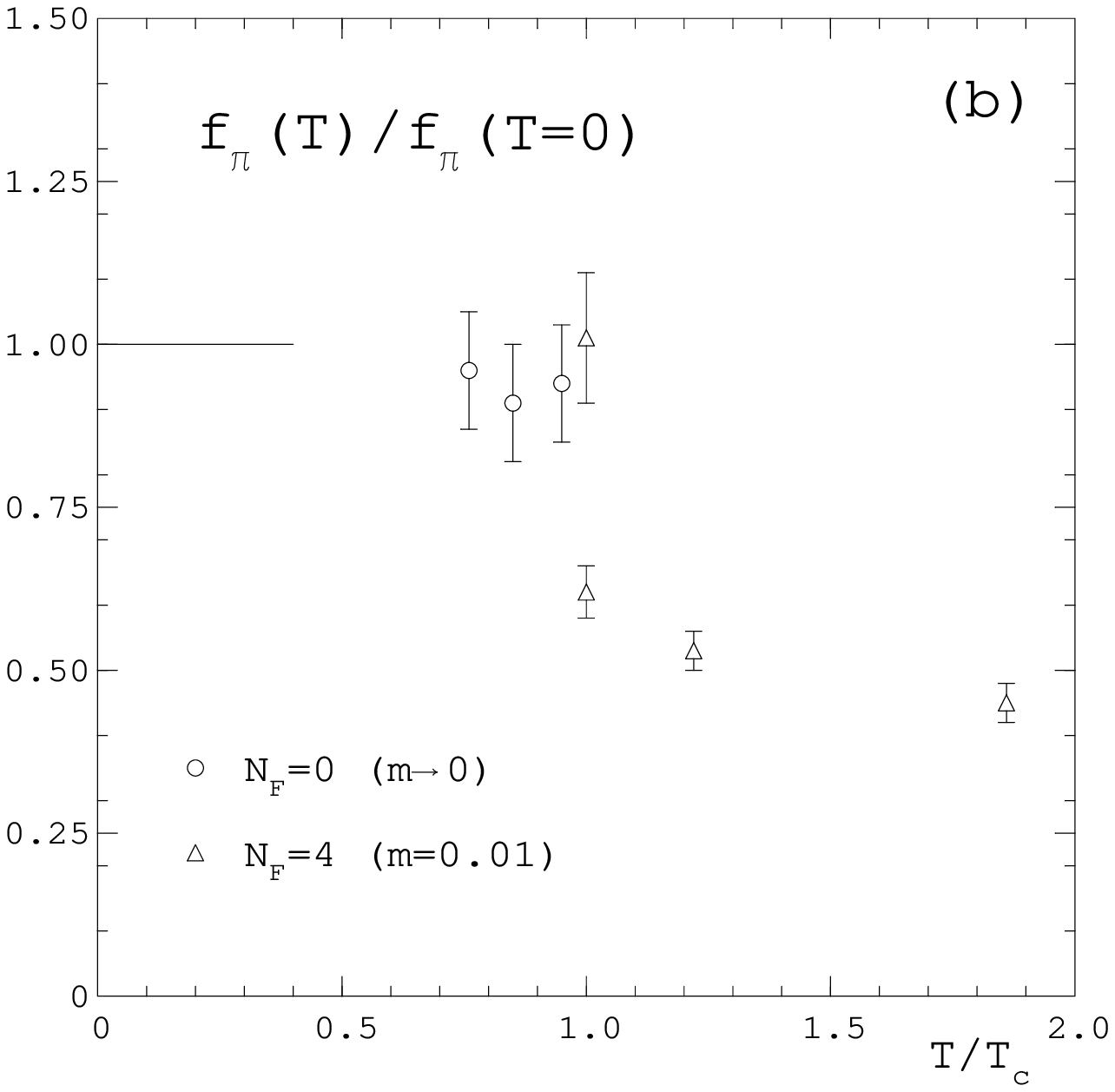,height=70mm,angle=-90}
   \end{center}
\end{minipage}
\caption{
Test of the GMOR relation at $T=0.92~T_c$ in quenched
QCD (a) and
the temperature dependence of the pion decay constant (b). In (a) we
compare the results obtained from the amplitude of the pion correlation
function through Eq.(15) (circles) and extrapolated to zero quark mass
with
the result obtained from the GMOR relation (star, on top of the circle at
$m_q a=0$). Also shown is the zero temperature result at this value of
the gauge coupling (square).}

\label{fig:gmor}
\end{figure}

The square of the matrix element appearing on the right hand side of
Eq.~\ref{fpmp} is proportional to the amplitude of the pion correlation
function and can be determined in a Monte Carlo calculation.
A comparison of $f_\pi$ determined this way with the ratio
$\langle \bar\psi \psi \rangle_{m_q=0} / a_\pi$ thus provides a direct
test of the GMOR relation at finite temperature. This is shown in
Fig.~\ref{fig:gmor}a at $T\simeq 0.92T_c$. Moreover, we can compare the
value of $f_\pi$ calculated at finite temperature with corresponding
zero temperature results. This is shown in Fig.~\ref{fig:gmor}b.
Clearly we do not have
any evidence for violations of the GMOR relation nor for a significant
change of $f_\pi$ with temperature below $T_c$. The sudden change above
$T_c$ reflects the drastic change in the structure of the pseudo-scalar
correlation function. It does no longer have a pole corresponding to a
Goldstone-particle (pion). The existence of a pseudo-scalar bound state above
$T_c$ thus is questionable. Certainly for large temperatures such a state
does not exist, the correlation function is dominated by a quark/anti-quark
cut. The notion of $f_\pi$ used for the square root of the amplitude of
the correlation function should thus be used with caution above $T_c$.

\begin{figure}[htb]
\begin{minipage}[t]{80mm}
  \begin{center}
\vskip -0.6truecm
    \leavevmode
    \epsfig{bbllx=100,bblly=245,bburx=450,bbury=600,
        file=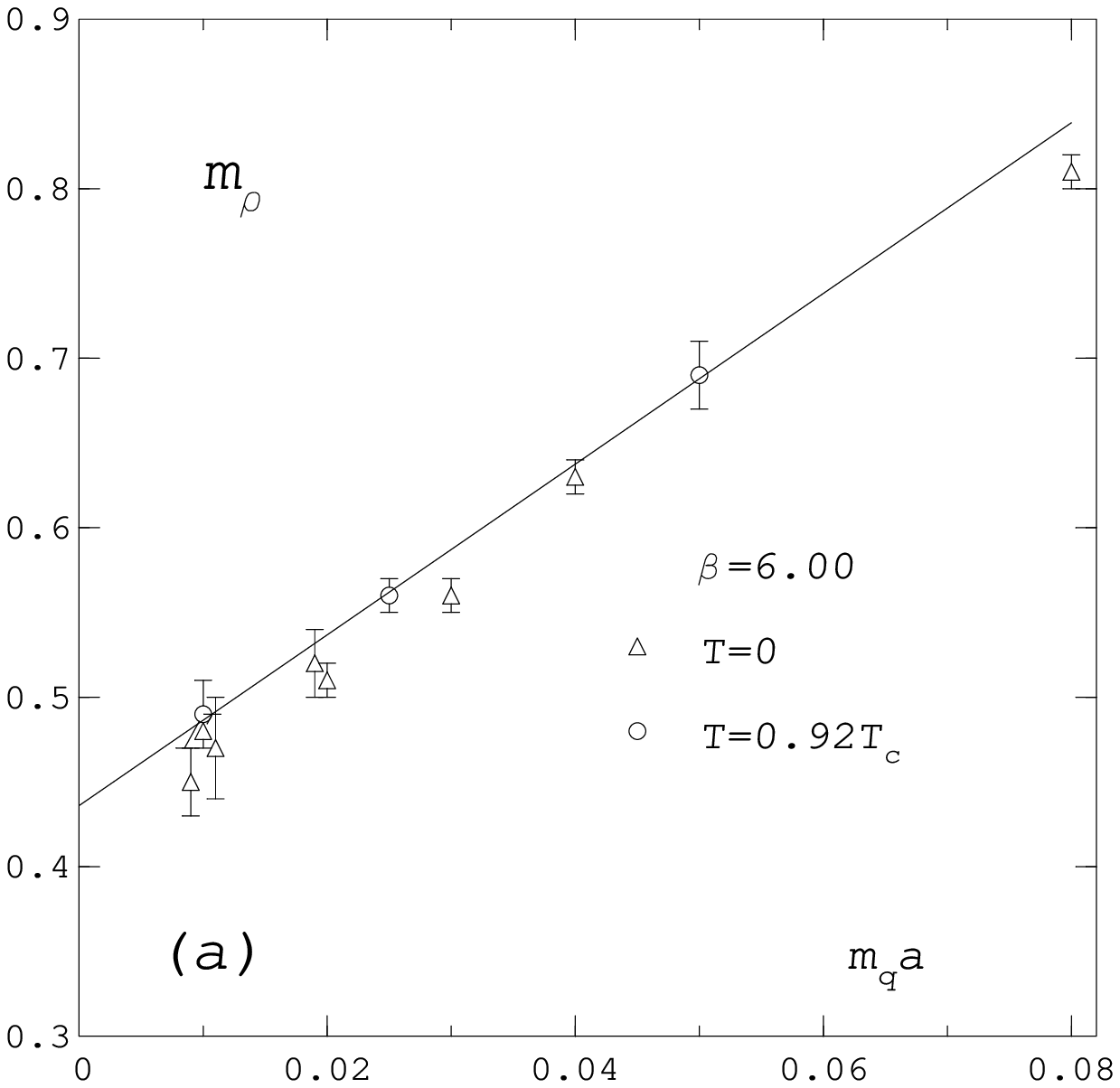,height=70mm,angle=-90}
   \end{center}
\end{minipage}
%
%
\begin{minipage}[t]{75mm}
  \begin{center}
\vskip -0.6truecm
    \leavevmode
    \epsfig{bbllx=100,bblly=245,bburx=450,bbury=600,
        file=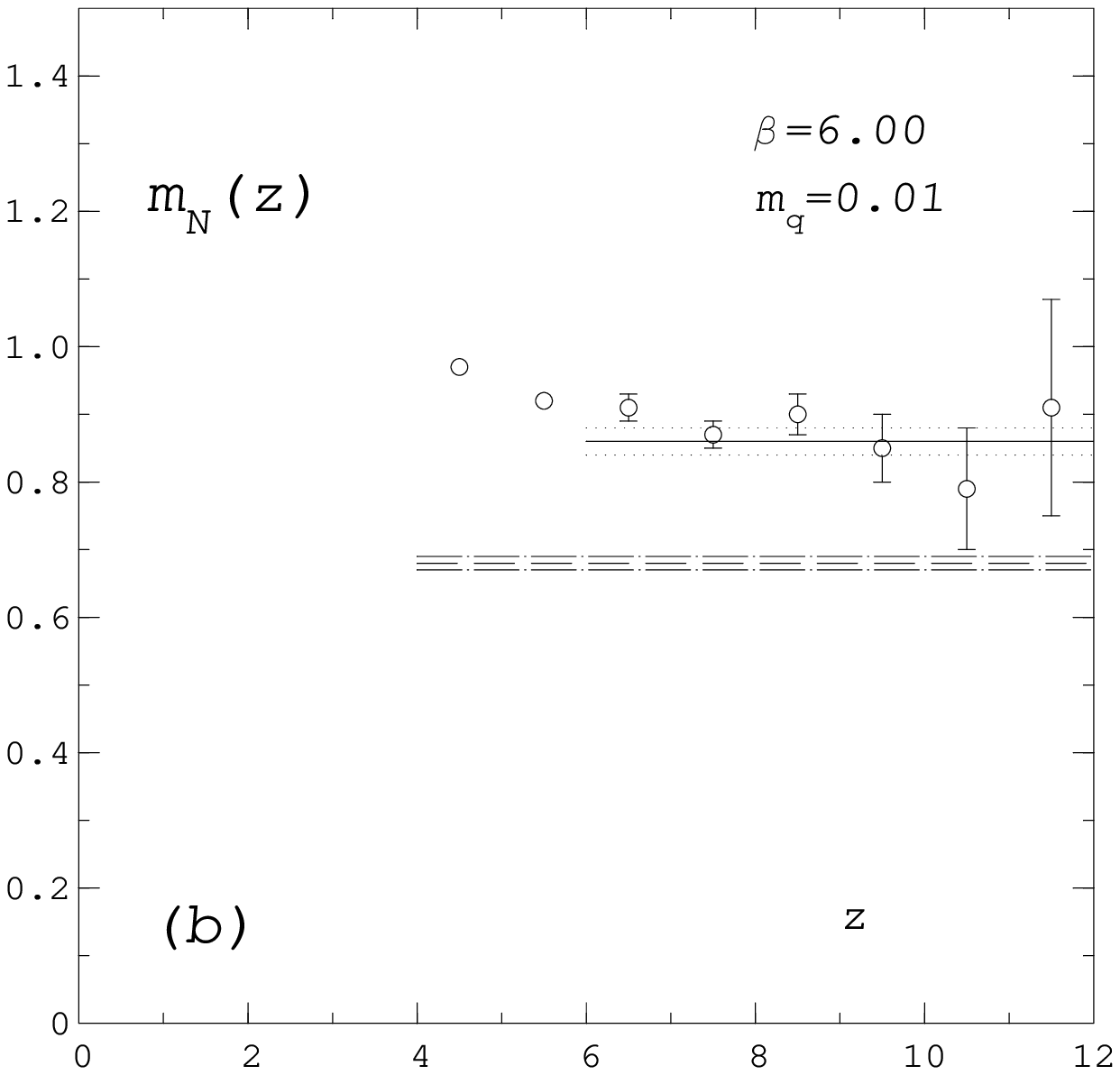,height=70mm,angle=-90}
   \end{center}
\end{minipage}
\vskip 0.5truecm
\caption{
The $\rho$-meson mass in quenched QCD for various values of the quark mass
(a) and local masses from the nucleon correlation function (b). The
simulations have been performed at fixed value of the gauge coupling,
$6 /g^2$=6.0, on a $32^3\times$8
lattice corresponding to $T\simeq0.92~T_c$ and
on various low temperature lattices [5].
In (b) we show estimates for
the nucleon mass obtained from the nucleon correlation function at distance
$z$. These local masses converge to the nucleon mass in the
limit $z\rightarrow \infty$. The horizontal band indicates the
corresponding extrapolation at zero temperature.
No temperature dependence is visible in the $\rho$-meson mass as well as
the nucleon mass (see text for appropriate subtraction of the finite
temperature Matsubara mode).}

\label{fig:rho}
\end{figure}

\vskip 10pt
\noindent
{\large \bf III.3.2 \hskip 0.3truecm  Vector meson mass and nucleon mass
\hfil}
\vskip 5pt
\noindent
The possibility of a variation of the $\rho$ meson mass with temperature
has been discussed a lot as this might lead to modifications of
dilepton spectra, which are experimentally detectable.
The temperature dependence of the meson masses has been discussed within
the framework of the OPE. Arguments have been given that the meson masses
are temperature independent up to $O(T^2)$ \cite{Ele93}.
The Monte Carlo calculations of the vector meson correlation function at
finite temperature also show no significant temperature dependence  of the
mass even close to $T_c$. In Fig.~\ref{fig:rho} we show the result of our
calculation of the screening mass in the vector channel correlation
function at $T\simeq 0.92T_c$ and compare this with zero temperature
calculations at the same value of the gauge coupling. There is no
evidence for any temperature dependence. The same holds true for the
nucleon mass, although the details are more subtle in this case. As can be
seen in Fig.~\ref{fig:rho}b there is a clear difference between the
local masses, $m_N(z) \sim \ln G_N(z)/G_N(z+1)$,
extracted on a large zero temperature lattice and those extracted at
finite temperature on a lattice of size $32^3\times 8$. However, as the
nucleon is a fermion, there is a non-negligible contribution from the
non-zero Matsubara mode to the nucleon screening mass,
\beqn
m_N =\sqrt{\tilde m_N^2 +(\pi T)^2}
\label{mn}
\eqn
After removing the thermal contribution, $\pi T$, the nucleon mass,
$\tilde m_N$, agrees with the zero temperature result within
statistical errors.

\vskip 10pt
\noindent
{\large \bf IV. \hskip 0.3truecm  Conclusions \hfil}
\vskip 5pt
\noindent
We have discussed recent detailed studies of the equation of state
of the $SU(3)$ gauge theory. Deviations from an ideal gas behaviour
are still significant at temperatures of a few times $T_c$.  
The now existing extrapolation of the $SU(3)$ equation of state
to the continuum limit does provide a good basis for quantitative comparisons
with perturbative calculations and phenomenological models \cite{models}. 

The discussion of basic properties of the thermodynamics of 
two-flavour QCD shows that there are strong
indications for a second order phase transition.
However, whether two-flavour QCD is, indeed, in the same universality class
as the $3d$, $O(4)$ model cannot yet be decided on the basis of the
existing numerical data and needs further detailed investigations on larger
lattices and with smaller quark masses.

At least up to temperatures of about $0.9T_c$ so far no  
statistically significant temperature dependence could be observed 
for basic properties of hadrons.
This is true for the chiral condensate, 
hadron masses as well as decay constants calculated in quenched QCD.
At least for the chiral condensate one expects, however, a strong
variation with temperature for $0.9T_c < T < T_c$, in particular in the 
case of a second order phase transition. This is clearly supported by
the currently available lattice calculations in two-flavour QCD. 
The influence of this on the behaviour of hadronic
parameters will certainly also be investigated in the near future.

\vglue 12pt
\noindent
{\large \bf Acknowledgements \hfil}
\vglue 5pt
The numerical work that has been reviewed here has been performed on the
Cray YMP at the HLRZ-J\"ulich and the QUADRICS parallel
computer at the University of Bielefeld, funded by DFG under contract No.
Pe 340/6-1. The work has also been supported by the Deutsche
Forschungs\-gemeinschaft under grant Pe 340/3-2 and Pe 340/3-3.

\end{document}